# ON TIME. VIB: QUANTUM-MECHANICAL TIME*

C. K. Raju

*Indian Institute of Advanced Study, Rashtrapati Nivas, Shimla 171 005*

***Prefatory Note**: This paper first appeared in *Physics Education* (India) **10**, 1993, 143–161, as Part 6b, or the 9th in a series of 10 papers 'On Time'. These papers later also appeared as chapters in a book (*Time: Towards a Consistent Theory*, Kluwer Academic, Dordrecht, 1994). This is here being reproduced verbatim, except for this prefatory note. A key aim of this series/book was to explain how both relativity and key aspects of quantum mechanics follow from a better understanding of the nature of time in classical physics. This prefatory note summarizes some subsequent developments, and clarifies some key points.

The earlier papers in the series had explained the advantages of dispensing with the field in classical electrodynamics, and working directly with the functional differential equations (FDEs) that arise in the particle picture. An earlier paper/chapter ('On Time. 5b: Electromagnetic Time', *Physics Education* **9**, 1992, 251–265) had also explained the elementary mathematical theory of FDEs which makes them so radically different from the 'Newtonian paradigm' of ordinary differential equations (ODEs). The present paper explains how the peculiar features of mixed-type FDEs can be used to understand the most puzzling aspect of quantum mechanics, viz. quantum interference. The earlier published version is here being reproduced verbatim, except for this note.

The key difference between this paper and the corresponding book chapter is that the book includes the proof of the main theorem, as an appendix. A subsequent paper (C. K. Raju, *Found. Phys.* **34**, 2004, 937–62, arxiv.org:0511235) connected retarded FDEs to quantum mechanics at a more intuitive level, by numerically calculating the solution of the retarded 2-body FDEs for the classical hydrogen atom.

More recently, the connection of FDEs to quantum mechanics, which was reported here as a *theorem,* and sought to be established by *calculation* in Raju (2004), was subsequently advanced as a *conjecture* by M. Atiyah in his Einstein lecture of 21 October 2005, at the University of Nebraska-Lincoln, and in a subsequent lecture at the Kavli Institute of Theoretical Physics (http://online.kitp.ucsb.edu/online/strings05/atiyah/). This author's prior work connecting FDEs to quantum mechanics went unacknowledged, and remained unacknowledged in a subsequent article (*Notices of the AMS*, **53,** 2006, 674–78, http://www.ams.org/notices/200606/comm-walker.pdf), which reiterated Atiyah's priority.

This raised some peculiar ethical issues since Atiyah had already been explicitly informed of this author's work within a few days of his Einstein lecture, and had subsequently also responded to the author (M. Atiyah, personal communication). Also, an author of that *Notices* article, reporting Atiyah's Einstein lecture, confirmed the natural assumption that the article was shown to Atiyah before submission (M. Walker, personal communication). Setting aside the ethical issues, which have been discussed elsewhere, and also partly investigated and reported on its website by the Society for Scientific Values (http://scientificvalues.org/cases.html), there are a couple of key point about physics here.

The first concerns the mistake evident from the very terminology of "Atiyah's hypothesis": that physics should use FDEs instead of ODEs or partial differential equations (PDEs), *as a hypothesis*. As this author has repeatedly pointed out earlier, using FDEs does *not* require any new hypothesis. The 2004 paper, cited above, specifically clarified that FDEs are equivalent to a *coupled* system of PDEs and ODEs, and arise naturally in physics. For example, in the many-body problem of electrodynamics, the equations of particle motion are ODEs (corresponding to the Heaviside-Lorentz force) which couple to PDEs (Maxwell's equations) for the electromagnetic field. This coupled system is equivalent to the FDEs of the pure particle picture.

The other issue of concern here is the *type* of the FDEs. Retarded FDEs already destroy the Newtonian paradigm by bringing in history-dependence. This can indeed explain *some* puzzling aspects of quantum mechanics, as explicitly argued in this author's 2004 paper, cited above, which solved the retarded FDEs for the hydrogen atom. However, a natural explanation of quantum interference requires *mixed-type* FDEs which are here assumed to obtain the requisite structure of time. It is then argued that the (temporal) logic corresponding to this structure of time is a quantum logic. A parallel computer is a concrete desktop model to help understand this logic. However, with a view to be rigorous, this paper presented this simple argument at the abstract level of axiomatic quantum mechanics.

It is assumed that the reader knows that the picture of unitary evolution in Hilbert space can also be used for classical statistical mechanics. Secondly, although the structured-time interpretation of quantum mechanics (this paper) has almost nothing in common with this author's earlier interpretation of q.m. (*Int. J. Theor. Phys.* **20,** 1981, 681–96), the earlier paper connected the Schrödinger equation (unitary evolution in Hilbert space) to equilibrium (indifference to choice of time origin), using an argument from the theory of stationary stochastic processes. Since that argument was so general, it could be reused.

However, that does not explain non-commutativity, or, equivalently, non-existence of joint probability distributions, or quantum interference. **It is against this background that the paper supposes that the key aspect of quantum mechanics which requires explanation is , in mathematical terms, the non-existence of joint probability distributions.**

As regards measurement, or the collapse postulate, the structured time interpretation (STI) already has an explicit underlying dynamics using mixed-type functional differential equations. (This dynamics relates to classical *electrodynamics*, which should not be confounded with classical *mechanics* or the 'Newtonian paradigm' of ordinary differential equations, which is rejected by the STI. Accordingly, the STI, though non-local, is NOT a hidden variable theory.) The dynamics underlying the STI applies to both equilibrium (unitary evolution) and non-equilibrium situations. Hence, there is no specific need of a separate measurement postulate, though the STI naturally assumes that a dynamical variable can be measured only when it has a definite value.

The basic assumption of mixed-type FDEs, here called the hypothesis of a 'tilt in the arrow of time', was earlier related to the empirical existence of small amounts of advanced electromagnetic radiation. This, incidentally, was explicitly predicted by this author's version of the absorber theory of radiation (*J. Phys. A: Math. Gen.* **13**, 1980, 3303–17). However, as clarified by a later chapter of the book cited above (*Time: Towards a Consistent Theory*, p. 227), there is another way of looking at the matter: 'The proposal for a tilt in the arrow of time only carries the relativistic postulate to its logical conclusion…. Strictly speaking the proposal does not even involve any new hypothesis; rather it discards the unsound, though traditional, hypothesis of "causality" or perfect time asymmetry.' This way of looking at things was also noted in a review of the book (*Foundations of Physics*, **26**, 1996, 1725–31), which quoted Bohm to the effect that ' "Progress in science is usually made by *dropping* assumptions" ' and added that 'Raju…has…seen that the assumption that needs to be dropped is common-sense causality.'

ABSTRACT. We present a brief exposition of the orthodox axiomatic approach to q.m., indicating the relation to the text-book approach. We explain why the usual axioms force a change of logic. We then explain the attempts to *derive* the Hilbert space and the probability interpretation from a new type of 'and' and 'or' or a new type of 'if' and 'not'. Included are the Birkhoff-von Neumann, Jauch-Piron, and quantum logic approaches, together with an account of their physical and mathematical obscurities.

Instead of entering the labyrinth of subsequent developments, which seek new algebraic structures while accepting the old physical motivation, we present an exposition of the structured-time interpretation of q.m., which seeks a new physical motivation.

We saw in Part VB[*] that, with a tilt in the arrow of time, the solutions of the many-body equations of motion are intrinsically non-unique. In Part VIA we had indicated how this non-uniqueness relates to a change in the logic of time. We now explain how the resulting changes in the logic and structure of time lead to a new type of 'if' and 'not', of the kind required by q.m., while escaping from the criticism which applies to the earlier 'quantum logic' approaches.

We briefly indicate the analogy between this logic and the temporal logic required for the formal semantics of parallel-processing languages like OCCAM, and distinguish the structured-time interpretation from the superficially similar many-worlds interpretation and the transactional interpretation of q.m.

## 1  Introduction

**T**HE preceding part introduced the problem of a non-trivial structure of time: the (local) topology of time, in the real world, might be different from that of the real line. The real-line topology differs from the mundane view of a past-linear future-branching time, used to demarcate and validate physics. Moreover, there is possible incoherence about the structure of time, even within physics, as different structures may be simultaneously implicit.

We explained how the notion of a structure of time could be formalized in terms of properties of the earlier-later relation (U-calculus) or, more generally, using an appropriate (temporal) logic.

This part deals with two earlier claims (a) that an appropriately structured time could be related to the change of logic required by the axiomatic formulation of q.m., and (b) that the hypothesis of a tilt in the arrow of time implies such an appropriate structure. The other consequence of the basic hypothesis, viz. non-locality, is hardly a serious drawback since we saw in Part VIA that locality is a fuzzy and metaphysical requirement which lacks a basis even in classical physics.

To reiterate, the aim of this part is to present an exposition of the structured-time interpretation of q.m. which relates the many-body equations of motion of non-local classical (relativistic) mechanics, the emergence of a logical structure, or a non-trivial topology of time,[1] and the mathematical formalism of q.m.

---

§ 2 presents an exposition of the orthodox Hilbert-space axiomatics of q.m. and relates it to the usual textbook approach. § 3 explains why the orthodox axiomatic approach forces a change of logic and goes on to present an exposition of the 'quantum logic' approach, its relation to the Hilbert-space axiomatics, and its obscurities. The idea is to distill the body of q.m. to its algebraic and logical 'skeleton'. Finally, § 4 presents the structured-time approach, relating a structure of time to the 'quantum logic' approach. The programme is to reconstruct the algebraic skeleton from the hypothesis of a tilt in the arrow of time (presented in Part VB). It remains to be seen whether this procedure generates the body of q.m.!

Since the structured-time approach involves inputs from diverse areas at the frontiers of knowledge (such as the many-body problem, the theory of counterfactuals and conditionals in general, parallel processing, and temporal logic), an in-depth understanding may require some mathematical sophistication. But I believe the basic ideas may be grasped by anyone who has some familiarity with the modern (Lebesgue) integral and Hilbert spaces. To relate this approach to conventional q.m., we need the connections established in § 2 and § 3 below, for which there is unfortunately no convenient reference.

## 2  The orthodox formalism of q.m.

The chief features of the orthodox formalism of quantum mechanics are the following. For those who skip this section: the main point is to get rid of the usual hang-up with $|\psi|^2$ as a probability density, and to introduce some form of the projection postulate to work with.

(i) **State space**: An abstract separable Hilbert space **H** to which the state vectors of a system belong.

(ii) **Operator representation**: A correspondence between classical dynamical variables and densely defined (and maximally extended) self-adjoint operators on **H**, subject to the restriction that canonically conjugate variables satisfy the commutation relations

$$[\hat{p}, \hat{q}] = i\hbar, \qquad (1)$$

where $\hbar$ denotes the Planck constant divided by $2\pi$, the bracket denotes the commutator, and the operator on the right is a scalar multiple of the identity. In the textbook approach, in the configuration-space representation, one takes $\hat{q}$ to be the multiplication operator, $\hat{q}: \psi(q) \to q\psi(q)$, and $\hat{p}$ to be the differentiation operator, $\hat{p}: \psi(q) \to i\hbar(\partial\psi/\partial q)$ (with appropriate boundary conditions).

(iii) **The probability interpretation**: The probability interpretation is closely related to the operator representation, a fact sometimes obscured by texts. The relation is obtained by means of the spectral theorem: an observable or a self-adjoint operator $T$ generates a 'resolution of the identity', or a spectral measure or a projection-valued measure $E$ on **H**, such that

$$T = \int_{\sigma(T)} \lambda \, dE(\lambda), \qquad (2)$$

where $\sigma(T)$, the spectrum of $T$, is a subset of the real line $R$. In general, if $\hat{p}$ is a dynamical variable, and $E_p$ is the corresponding spectral measure, the probability that $p$ lies in the Borel subset[2] $A$ of $R$ is

$$Pr(p \in A) = <E_p(A)\psi, \psi>, \qquad (3)$$

where $<\cdot, \cdot>$ denotes inner product in **H**, and $\psi$ is the state with $|\psi| = 1$. For an arbitrary $\psi$, $<E_p(\cdot)\psi, \psi>$ is a regular Borel measure, which is a probability measure when the state is normalized.

The correspondence with the textbook approach is obtained as follows. In the first place, the form of the operator (whether $\hat{q}$ above or $i\hbar\, \partial/\partial p$) is unimportant. A self-adjoint operator $T$ is really characterized by giving its spectrum $\sigma(T)$, together with multiplicities. For an observable $T$, the spectrum $\sigma(T) \subseteq R$ and corresponds to measurable values.

For, say, the position operator $\hat{q}$, if we believe that the spectrum $\sigma(\hat{q}) = R$, with no multiplicity, the spectral multiplicity theorem[3] (Hahn-Hellinger theorem) allows us to recover the usual configuration-space representation. The theorem provides a unitary map, between the abstract Hilbert space **H** and $L_2(\sigma(\hat{q})) \equiv L_2(R)$, which carries $\hat{q}$ to the multiplication operator on $L_2(R)$, and the spectral measure $E_q$ to the spectral measure $E$ on $L_2(R)$ corresponding to multiplication by characteristic functions. Thus,

$$Pr(q \in A) = <E_q(A)\varphi, \varphi> = <\chi_A \psi, \psi> = \int_R \chi_A \psi\overline{\psi} = \int_A |\psi|^2 \qquad (4)$$

recovers the more usual form of the probability interpretation. The uncertainty principle is an easy consequence of non-commutativity (1) and the Schwartz inequality.

(iv) **Schrödinger equation**: The Schrödinger equation describes unitary evolution in this Hilbert space

$$\psi(t) = U(t)\psi(0). \qquad (5)$$

By Stone's theorem, any such (strongly continuous) one-parameter unitary group may be written as

$$U(t) = e^{-iHt}, \qquad (6)$$

where $H$ is self-adjoint. Hence, the infinitesimal form of (5) reads:

$$\frac{\partial \psi}{\partial t} = -i H \psi, \qquad (7)$$

which is the more usual form of the Schrödinger equation.

Physically, the infinitesimal generator $H$ is identified with the Hamiltonian, or energy operator. It is an extraordinarily curious fact that, modulo commutativity, the quantum Hamiltonian is the *same* function of the canonical variables as the classical Hamiltonian (when the latter exists).

(v) **The projection postulate**: The naive formulation of the projection postulate, for the paradigmatic observable with discrete spectrum, is the following.[4]
'…any result of a measurement of a real dynamical variable is one of its eigenvalues…,
…if the measurement of the observable *x* for the system in the state corresponding to $|x>$ is made a large number of times, the average of all the results obtained will be $<x|\xi|x>$…,
…a measurement always causes the system to jump into an eigenstate of the dynamical variable that is being measured….'

The result of measuring such an observable always is an eigenvalue. As a 'consequence' of the measurement process, the system is thrown (discontinuously) into the eigenstate corresponding to the measured eigenvalue. If the same measurement is repeated immediately, the same eigenvalue results. Von Neumann incorrectly[5] supposed that the postulate could be generalized in a straightforward way to observables with continuous spectra.

The precise formulation, even for an observable with discrete spectrum, is messy. One must account for incomplete measurements, possible degeneracy (non-zero multiplicity in the spectrum), and superpositions. The last can be achieved by viewing states as density matrices (trace-class operators, provided by Gleason's theorem, normalized to trace unity; this theorem was described in Part VIA).

Now, if *T* is an observable with discrete spectrum, the integral in (2) reduces to a sum,

$$T = \sum_i \lambda_i P_i, \qquad (8)$$

where $\lambda_i$ are the eigenvalues and $P_i$ are the corresponding eigenprojections. Let $A \in \boldsymbol{B_R}$ (*A* is a Borel subset of *R*). The act of measurement *conditioned on the statement T ∈ A* (partial measurement) transforms the state ρ to the (non-normalized) state ρ′,

$$\rho' = \sum_{\lambda_i \in A} P_i \rho P_i, \qquad (9)$$

the sum being taken over those indices *i* for which $\lambda_i \in A$. In the case where *A* is a singleton, $A = \{\overline{\lambda}\}$, (9) reduces to a complete measurement, and the system is thrown into an eigenstate. This process does not work if the observable is degenerate.

## 3  From quantum logic to the formalism of q.m.

*3. 1  Non-commutativity and non-existence of joint distributions*

The relationship of the axiomatic approach to the textbook approach may be clear, but the relation to phenomena remains a mystery. Understanding this mystery may require a long process of distillation.

What are the chief new features of this formalism? We have already encountered the Hilbert space and the picture of unitary evolution in the context of classical statistical mechanics (Part IV, Box 3). One could extend this picture to represent dynamical variables by self-adjoint operators.

The new feature however is non-commutativity: the approach of classical statistical mechanics always results in a commuting algebra of observables. Since non-commuting operators cannot be simultaneously diagonalized, this non-commutativity gives rise to a peculiar difference between classical and quantum probabilities: *a joint probability distribution does not exist* for canonically conjugate (non-commuting) dynamical variables, as observed[6] and later proved by Wigner.[7] The chief problem then would seem[8] to be the explanation of the origin of these peculiar quantum probabilities.

### 3.2 *Need for a change of logic: failure of the distributive law*

Now probabilities may be defined on a $\sigma$-algebra $M$ of subsets of a given set $X$, using $\cup$, $\cap$ and $\subseteq$, or on a logic of sentences. In the latter setting, the usual measure-theoretic approach to probability is recovered by identifying the usual set-theoretic operations with the logical operations required to define them: '*not*' with complement $'$, '*and*' with $\cap$, '*or*' with $\cup$, and $\Rightarrow$ with $\subseteq$. The usual calculus of sentences results in a Boolean algebra isomorphic to the algebra of subsets of a given set.[9] In the Birkhoff-von Neumann (BN) approach[10] the peculiarities of quantum probabilities are explained by asserting that the logic of q.m. differs from classical logic in that the *distributive law* between '*and*' and '*or*' *fails*. In a double-slit experiment, to say that 'the electron reached the screen *and* passed through slit A *or* slit B', is *not* the same as saying that 'the electron reached the screen *and* passed through slit A *or* the electron reached the screen *and* passed through slit B'. In one case one gets a diffraction pattern, in the other case a superposition of two Gaussians.

The failure of the distributive law means that a joint probability distribution cannot be defined; for example, the marginal distributions would fail to be additive.

$$Pr\{a \in A \ \& \ (b \in B \ or \ b \in C)\} \neq Pr\{a \in A \ \& \ b \in B\} + Pr\{a \in A \ \& \ b \in C\}, \qquad (10)$$

even if the '*or*' is exclusive, i.e., $B$ and $C$ are disjoint.

The BN approach, therefore, advocates a change of the logic on which the probabilities are defined. Probabilities, such as those on the left hand side of (4), are defined on sentences, but the '*and*' and '*or*' used to compound these sentences are such that the distributive law fails. One therefore obtains a more general algebraic structure, rather than the usual Boolean algebra (or $\sigma$-algebra), on which probabilities are to be defined, as countably additive, positive functionals with total mass 1.

### 3.3 *Birkhoff-von Neumann approach and the orthodox formalism of q.m.*

*3.3.1 The lattice of projections.* The BN approach begins by noticing that the subspaces[11] of a Hilbert space form a lattice, or a 'logic', with the desired properties. One may identify a subspace of a Hilbert space with the orthogonal projection onto that subspace. We now define an '*and*' ($\wedge$) and '*or*' ($\vee$) as follows. If $P_1$ and $P_2$ are two orthogonal projections on the subspaces $R(P_1)$ and $R(P_2)$ respectively, then $P_1 \wedge P_2$ is the projection on the subspace $R(P_1) \cap R(P_2)$, while $P_1 \vee P_2$ is the projection on the smallest subspace which contains both $R(P_1)$ and $R(P_2)$.

If the dimension of the Hilbert space is ≥2, and $P_1$, $P_2$, $P_3$ are taken, as in Fig. 1, as the projections on the *x*-axis, *y*-axis, and the line *y*=*x*, then it is clear that the distributive law fails: $P_1 \wedge (P_2 \vee P_3) = P_1$, but $P_1 \wedge P_2 = 0$, $P_1 \wedge P_3 = 0$, and $0 \vee 0 = 0$, so that $P_1 \wedge (P_2 \vee P_3) \neq (P_1 \wedge P_2) \vee (P_1 \vee P_3)$. Conversely, the algebraic structure corresponding to the usual sentence calculus is a *distributive* lattice or a Boolean algebra. Thus, the Hilbert space is related naturally to the failure of the distributive law.

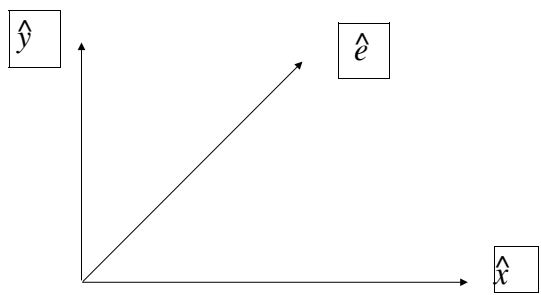

**Fig. 1: Failure of the distributive law**

*If the projections $P_1$, $P_2$, $P_3$ are defined by $P_1 \mathbf{a} = \mathbf{a}.\hat{x}$, $P_2 \mathbf{a} = \mathbf{a}.\hat{y}$, $P_3 \mathbf{a} = \mathbf{a}.\hat{e}$, for any vector $\mathbf{a}$, then the join of any two of these is the projection on the plane, while their meet is zero. Hence the join of any two meets is zero, and cannot equal the meet of any one projection with the join of the other two.*

*3. 3. 2 Order relation and orthocomplement.* The axioms for the usual sentence calculus may be formulated in terms of '*not*', '*and*' (∧) and '*or*' (∨), or, more usually, in terms of '*not*' (~) and implication (⇒). In algebraic terms, following the algebraization of logic initiated by Boole, '*if*' and '*not*' may be described respectively by an *order relation*[12] ≤, and an *orthocomplement* ′. These may be used to describe the properties of the lattice of projections **P**: $P_1 \leq P_2$ exactly if $\mathbf{R}(P_1) \subseteq \mathbf{R}(P_2)$, i.e., the subspace onto which $P_1$ projects must be a subset of the subspace onto which $P_2$ projects. The orthocomplement $P'$ is the projection on the null space of $P$ denoted by $N(P)$.

Given an order relation, the '*and*' and '*or*' may be re-interpreted: $P_1 \wedge P_2$ is the greatest lower bound (g.l.b., infimum), while $P_1 \vee P_2$ is the least upper bound (l.u.b., supremum) of the two-element set $\{P_1, P_2\}$. One would expect de Morgan's laws to hold. One may also define the notion of *orthogonality*, $P \perp Q$ if $P \leq Q'$.

*3. 3. 3 Geometrical interpretation.* All these notions have simple geometrical meanings in 3-dimensional Euclidean space. The closed *subspaces* are: the point at the origin, lines through the origin (extended to infinity in both directions), planes through the origin, and the whole space. The *orthogonal projections* are precisely that: if $P$ is the projection onto a line or a plane, the result of applying $P$ to a vector is obtained by dropping a perpendicular to the line or the plane in question. The *partial order* is set-theoretic inclusion, and the infimum is the set-theoretic intersection. The l.u.b. of two lines is the plane they span. The *orthocomplement* of a line is the plane perpendicular to it. *Orthogonality* just means perpendicularity of the corresponding subspaces.

*3. 3. 4 Dynamical variables, random variables and self-adjoint operators.* Apart from the failure of the distributive law, what is the point of studying the lattice of projections? An immediate application is that it leads to the operator representation of observables (hence the probability interpretation) in a natural way.

> **Box 1. Collected Definitions**
>
> **Basics**
>
> A *poset* is a pair $(\mathbf{P}, \leq)$, where $\mathbf{P}$ is a set and $\leq$ is an *order relation*. That is, $\forall\, a, b, c \in \mathbf{P}$, (i) $a \leq a$ ($\leq$ is reflexive), (ii) $a \leq b$ and $b \leq c \Rightarrow a \leq c$ ($\leq$ is transitive), (iii) $a \leq b$ and $b \leq a \Rightarrow a = b$ ($\leq$ is anti-symmetric).
>
> In a poset $(\mathbf{P}, \leq)$, $a \vee b$ denotes the *supremum* of $\{a, b\}$. That is, (i) $a \leq a \vee b$, and $b \leq a \vee b$ ($a \vee b$ is an upper bound), (ii) if $c \in \mathbf{P}$ is such that $a \leq c$, and $b \leq c$, then $a \vee b \leq c$ ($a \vee b$ is the least upper bound). Similarly, $a \wedge b$ denotes the *infimum* or the greatest lower bound. These notions may be extended to sets with an arbitrary number of elements. 0 is the g.l.b. for $\mathbf{P}$, while 1 is the l.u.b. for $\mathbf{P}$. A poset is called bounded if 0 and 1 exist.
>
> *b covers a* if $a \neq b$, $a \leq b$, and $\forall\, x \in \mathbf{P}, a < x \leq b \Rightarrow a = b$. In a *Hasse diagram*, an element $b$ which covers $a$ is placed directly above $a$. An element of $\mathbf{P}$ which covers 0 is called an *atom*. P is called *atomic* if, for any $x \in \mathbf{P}$, there is an atom $a \leq x$.
>
> **Lattice**
>
> A *lattice* is a set L with binary operations $\vee, \wedge$, which satisfy, for all $a, b, c \in$ L,
> (i) (associative laws) $a \vee (b \vee c) = (a \vee b) \vee c, \quad a \wedge (b \wedge c) = (a \wedge b) \wedge c$,
> (ii) (commutative laws) $a \vee b = b \vee a, \quad a \wedge b = b \wedge a$,
> (iii) (absorption laws) $a \vee (a \wedge b) = a, \quad a \wedge (a \vee b) = a$,
> (iv) (idempotent laws) $a \vee a = a, \quad a \wedge a = a$
> [(iv) is a consequence of (i)-(iii)]. One may define an order relation in a lattice by
>
> $a \leq b \Leftrightarrow a = a \wedge b$, or, equivalently, $a \leq b \Leftrightarrow b = a \vee b$.
>
> In a *σ-lattice*, countable sets admit suprema and infima. In a *complete* lattice, all sets admit suprema and infima.
>
> **Orthogonality and orthocomplement**
>
> An *orthocomplementation* on a bounded poset is a unary operation ′ which satisfies, $\forall\, a, b$,
> (i) (antitone) $a \leq b \Rightarrow b' \leq a'$,
> (ii) (period 2) $a'' = a$,
> (iii) (orthogonal decomposition) $a \vee a' = 1, a \wedge a' = 0$.
> An *orthoposet* is a poset with an orthocomplementation. In an orthoposet, $a$ is *orthogonal* to $b$, written $a \perp b$, if $a \leq b'$. Orthogonality is a symmetric relation.

The first step is that classically an observable or a dynamical variable is a random variable. One observes the dynamical variable, and the observed values vary or show some dispersion or scatter.

Now, in the usual measure-theoretic approach to probabilities, a random variable is a measurable *function*. Given a set *X*, and a Boolean σ-algebra, *M*, of subsets of *X*, a (real-valued) random variable is a function $f: X \to R$, such that $f^{-1}(A) \in M$ whenever $A \in \boldsymbol{B_R}$.

**Distributive and modular laws**

Given a lattice L, for $a, b, c \in$ L, a triple $(a, b, c)$ is called a *distributive triple* if the distributive laws

$$(a \wedge b) \vee c = (a \vee c) \wedge (b \vee c), \quad (a \vee b) \wedge c = (a \wedge c) \vee (b \wedge c)$$

hold. A *distributive* ortholattice is a *Boolean algebra*. L is *modular* exactly if $a \leq b$ implies that $(a, b, x)$ is a distributive triple, for every $x \in$ L. L is *orthomodular* if $a \leq b$ implies that $(a, b, a')$ is a distributive triple. In general, we have the relations

$$\text{distributive law} \Rightarrow \text{modular law} \Rightarrow \text{orthomodular law},$$

so that a Boolean algebra is a modular ortholattice, which, in turn, is an orthomodular lattice.

In a lattice L, $\{a, b\}$ is called a *modular pair*, if $\forall x \in$ L, $x \leq b \Rightarrow (a, b, x)$ is a distributive triple. L is called *semi-modular* if whenever $\{a, b\}$ is a modular pair, so is $\{b, a\}$.

Modularity holds for the lattice **P(H)** of projections on a Hilbert space, when **H** is *finite-dimensional*, but fails otherwise since the sum M+N, of two subspaces, M, N, need not be closed. The *orthomodular law*,

$$a \leq b \Rightarrow a \vee (a' \wedge b) = b,$$

always holds for **P(H)**.

One can also think of other notions of modularity in terms of a generalized associative law. One has, for instance, the *modular law*

$$a \leq c \Rightarrow a \vee (b \wedge c) = (a \vee b) \wedge c.$$

Similarly, $\{a, b\}$ could be called a modular pair if $\forall x \in$ L, $x \leq b \Rightarrow (x \vee a) \wedge b = x \vee (a \wedge b)$.

Orthomodular lattices are characterized among ortholattices by the property that they have no subalgebra isomorphic to the ortholattice shown in the 'benzene ring' Hasse diagram (Fig. 2).

On the above ortholattice, one could define a measure which would fail to be monotone: $\mu(0) = 0$, $\mu(a) = 2$, $\mu(b) = 1$, $\mu(x) = \infty$, for any other $x$. However, on an orthomodular lattice, a non-negative additive map is always monotone.

What one actually requires for this mysterious textbook definition is the *inverse* map, $f^{-1}: \boldsymbol{B_R} \to M$, which is an isomorphism between the two $\sigma$-algebras:

$$f^{-1}(\cup A_i) = \cup f^{-1}(A_i),$$

$$f^{-1}(\cap A_i) = \cap f^{-1}(A_i),$$

> **Compatibility**
>
> An orthomodular poset may contain a Boolean subalgebra. An example is the *proposition range* of a classical dynamical variable $m$, given by $\{p \equiv m \in E \mid E \in \boldsymbol{B_R}\}$. A *block* in an orthomodular lattice L is a maximal Boolean subalgebra. Two elements $x$ and $y$ *commute*, $x\,C\,y$, if $x$ and $y$ are contained in a block. Commutativity is a symmetric relationship in an orthomodular lattice. Equationally, the relationship of commutativity in an orthomodular lattice is defined as follows.
> $x\,C\,y$ if and only if any one of the following holds:
>
> $$x = (x \wedge y) \vee (x \wedge y'),$$
> $$x = (x \vee y) \wedge (x \vee y'),$$
> $$x \wedge (x' \vee y) = x \wedge y.$$
>
> The *commutator* $c(x, y)$ of $x, y$ is defined by
>
> $$c(x, y) = (x \wedge y) \vee (x \wedge y') \vee (x' \wedge y) \vee (x' \wedge y').$$
>
> $x\,C\,y$ if and only if $c(x, y) = 1$. The *centre* C(L) of an orthomodular lattice L is the set of those elements which commute with every other element:
>
> $$C(L) = \{z \in L, z\,C\,x,\ \forall\, x \in L\}.$$
>
> C(L) is a Boolean subalgebra of L. An orthomodular lattice is called *irreducible* if the centre is trivial: $C(L) = \{0, 1\}$.
>
> **Hilbert lattice**
>
> The lattice L satisfies the *exchange axiom*, if whenever $a$ covers $a \wedge b$, then $a \vee b$ covers $b$. A complete, atomic, orthomodular, infinite-dimensional lattice, satisfying the exchange axiom, is called a *Hilbert lattice*. A Hilbert lattice does not necessarily arise as the lattice of closed subspaces of a Hilbert space (real, complex, or quaternionic).

$$f^{-1}(A^c) = (f^{-1}(A))^c, \qquad (11)$$

where $i$ will always denote an index running over finite or countable values.

When the setting is changed from a $\sigma$-algebra to the lattice $\mathbf{P}(\mathbf{H})$ of projections on a Hilbert space $\mathbf{H}$, it is, therefore, natural to define a random variable in this way as an isomorphism $m: \boldsymbol{B_R} \to \mathbf{P}(\mathbf{H})$ into a Boolean subalgebra of $\mathbf{P}(\mathbf{H})$:

$$m(\cup A_i) = \vee m(A_i),$$

$$m(\cap A_i) = \wedge m(A_i),$$

$$m(A^c) = (m(A))'. \tag{12}$$

The definition (12) makes sense for a broader class of algebraic structures. But if the basic algebraic structure is the lattice of projections on a Hilbert space, then (12) means that an observable, or a random variable, is automatically a projection-valued measure or a self-adjoint operator.

Similarly, one can define a state, or a probability measure on the lattice of projections. The text-book definition of a (positive) measure is the following. Given $(X, M)$ as above, a measure $\mu$ is a function $\mu : M \to R^+$ such that (i) $\mu(\phi)= 0$, (ii) if $A_i$'s are pairwise disjoint, $\mu\left(\cup A_i\right) = \sum \mu(A_i)$. A probability measure satisfies $\mu(X)= 1$.

In the lattice of projections, the notion of disjointness is replaced by orthogonality: if $P \perp Q$ then $P \wedge Q = 0$. So, one uses the conditions that (i) $\mu(0) = 0$, (ii) $\mu(\vee P_i) = \sum \mu(P_i)$, if the $P_i$'s are pairwise orthogonal. A probability measure satisfies $\mu(1) = 1$. Gleason's theorem now recovers the usual density matrix approach.

### 3. 4  The quantum logic approach

*3. 4. 1 The minimum requirements.* To summarize the preceding section, the crucial distinguishing feature of quantum probabilities is the *non-existence of joint distributions*. In the BN approach, this is achieved by defining the probabilities on the *non-distributive* lattice of projections on a Hilbert space. The operator representation of dynamical variables (or the 'probability interpretation') emerges as a bonus in this approach.

But the mysterious Hilbert space is still in the background. Can one account for this too, starting only from a new type of '*and*' and '*or*' or a new type of '*if*' and '*not*'?

The quantum logic approach tries to whittle down, a little further, the mystery of quantum axiomatics. What is the minimum structure necessary to speak of probabilities without a joint distribution?

Suppose we are given a set **P**, together with an order relation $\leq$, and an orthocomplement $'$. Such a triple $(\mathbf{P}, \leq, ')$ is called an *orthoposet*. For the above definition of random variable and probability measure to go through, the right hand sides should make sense. The minimum requirement is that **P** should be $\sigma$-orthocomplete, i.e., every countable collection of mutually orthogonal elements of **P** should admit a supremum in **P**.

*3. 4. 2 Orthomodularity.* There is one more technical requirement: **P** should be orthomodular (see box). Orthomodularity is a weak form of the distributive law:
  distributivity $\Rightarrow$ modularity $\Rightarrow$ orthomodularity.
The lattice of projections on a Hilbert space **H** is orthomodular, but fails to be modular if **H** is infinite dimensional.

The point of orthomodularity is this: if **P** fails to be orthomodular, measures on **P** may fail to be monotone. That is, we could have[13] $a \leq b$, but $\mu(a) \geq \mu(b)$, which is unacceptable.

Thus, the minimum necessities for life seem to be that **P** should be an orthomodular, $\sigma$-orthocomplete orthoposet. Unfortunately, such an object need not admit any probability measures at all. If one further postulates the existence of 'enough' probability measures on

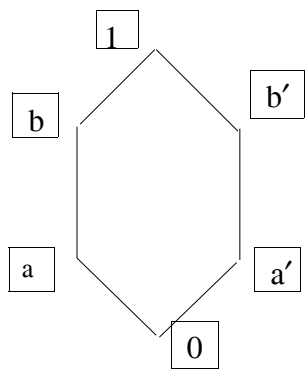

**Fig. 2: Example of a non-orthomodular lattice**

*A 'benzene ring' Hasse diagram provides an example of a non-orthomodular ortholattice. A measure defined on such a lattice could fail to be monotone. Hence the requirement of orthomodularity.*

*For the definitions and the notion of a Hasse diagram see Box 1: Collected definitions.*

**P**, or a 'full' or order-determining set of states on **P**, one obtains a *quantum-logic* proper, in the sense of Gudder. 'Enough' probability measures means that if $a \geq b$ then there is a probability measure $p$ such that $p(a) \geq p(b)$.

*3. 4. 3 Compatibility.* In the general setting of a quantum logic, there need be no operation corresponding to the usual algebraic *product* of two projections on a Hilbert space. So what happens to the relationship between non-commutativity and incompatibility or the non-existence of joint distributions?

One may recover this by *defining* commutativity or compatibility via the relation

$$x \, C \, y \iff y \equiv x \wedge (y \vee x'), \qquad (13)$$

which could be interpreted to mean: $x$ is compatible with $y$ if whenever $x$ is 'true', and $y \vee x'$ is 'true', then $y$ is 'true'.

In the lattice of projections, this notion of commutativity agrees with the usual notion. The relation to distributivity is this: the triple $\{a, b, c\}$ is distributive if any one is compatible with the other two. This provides one more reason to believe in orthomodularity — an orthoposet is orthomodular when the relation of compatibility is *symmetric.*

*3. 5 The Jauch-Piron approach*

In a quantum logic, the need to *postulate* a full set of states ('enough' probability measures) introduces a jarring note into the program of 'deriving' the Hilbert space from a new type of '*and*' and '*or*' or '*if*' and '*not*'. Can one give sufficient conditions on the algebraic structure so that this requirement is automatically satisfied?

An example of such sufficient conditions is provided by Piron's representation theorem. One needs to assume that the '*if*' and '*not*' give rise to
(i) an orthomodular lattice $L$, such that
(ii) $L$ is complete (suprema and infima exist for arbitrary sets),
(iii) $L$ is irreducible (0 and 1 are the only elements of $L$ which commute with all other elements), and
(iv) $L$ is semimodular ($x \leq y \implies y \wedge (y' \vee x) = x$).

Such an algebraic structure is called an irreducible CROC. Piron's representation theorem asserts that if $L$ is an *atomic*, irreducible CROC, then $L$ is isomorphic to the lattice of closed subspaces of a Hilbert space.

The significance of atomicity arises from the relation of atoms to Jauch-Piron states. A Jauch-Piron state is a subset $S \subseteq L$ satisfying

(i) $0 \notin S$, (ii) if $a \in S$ and $a \leq x$, then $x \in S$, (iii) if $a_i \in S$, then $\bigwedge_\iota a_i \in S$, (iv) $S$ is maximal with respect to the above properties.
Every Jauch-Piron state defines an atom, and vice versa.

### 3. 6 *Defects in the quantum logic approach*

Though the logico-algebraic approach initially seemed promising, it now seems riddled with physical and mathematical obscurities.

As one example of a mathematical difficulty, consider the following. The Piron representation theorem is one of the deeper results of the theory. But the Hilbert space provided by Piron's theorem is not necessarily the usual Hilbert space over the field of complex numbers: it is a Hilbert space over a 'division ring with involution', $F$. If this Hilbert space admits an 'observable' which is 'maximal' in some sense, one may show that $F$ is an extension of the reals. If one further assumes that the degree of this extension is finite, the Frobenius theorem[14] tells us that $F$ must be $R$ or $C$ (complex numbers) or the Quaternions. The final result is tantalizing, but what is one to make of the intermediate hypotheses?

As another example, consider the following. A lattice with all the nice properties of the lattice of projections is called a Hilbert lattice (see box). But a Hilbert lattice cannot necessarily be represented by the lattice of projections on a Hilbert space. Clearly there are limits to what can be achieved by a purely algebraic approach.

One could live with the mathematical difficulties, but the physical obscurities are fatal.

(i) The whole idea behind the approach is to motivate quantum axiomatics from a physical standpoint. Now we have already seen the emergence of some distinct structures
— Hilbert lattice (Birkhoff-von Neumann)
— quantum logic (Gudder)
— atomic irreducible CROC (Jauch-Piron)
The actual list is very much longer. How is one to choose between these structures?
For this one must go to the physical interpretation. The physical interpretation relates to the elements of the lattice: these are 'yes-no' type 'questions' (Jauch-Piron), or true-false 'outcomes of measurement' (Birkhoff-von Neumann). But one cannot hope to simultaneously measure incompatible observables, so the '*and*' does not make physical sense. *Hence one cannot, in principle, have a lattice*, and out goes the Hilbert space. One can form alternative, better motivated structures, but this does not lead to quantum physics as used by physicists for calculations.
(ii) While the algebraic approach sheds some light on the algebraic structures underlying classical mechanics and q.m., a problem common to all such approaches is the following. They push deeper into oblivion the link between classical and quantum dynamics, or the link between classical Hamiltonian evolution and evolution according to the Schrödinger equation.
(iii) This approach does not explain why the change of logic should be needed at the microphysical level, while usual Boolean logic is good enough at the macrophysical level, especially since the concepts of 'micro' and 'macro' tend to be anthropocentric. Going back to the 'clumsy fingers' kind of idea would defeat the whole purpose of the approach, irrespective of experiments on Bell's inequalities.

Thus, the logico-algebraic approach fails to explain the origin of the Hilbert space and the peculiar nature of quantum probabilities. The most sympathetic view that one can adopt, and the one that I will adopt in the sequel, is that this approach, while partly satisfying, needs further development, not so much at the level of formal axiomatic systems for quantum logics, but in terms of *a priori* motivation for the change of logic at the microphysical level. It is better to clarify that this is more or less the exact opposite of the typical approach in this area.

## 4  The structured-time interpretation of q.m.

### *4. 1  Motivation*

The structured-time interpretation[15] may be motivated as follows. That classical mechanics fails at the microphysical level is an established fact. But can one identify the intrinsic or *a priori* shortcomings of classical mechanics which led to this failure? Can one identify those microphysical shortcomings of classical mechanics which need to be remedied in order to obtain the *formalism* of q.m.? As an analogy, there was a theoretical problem with the measurement of time in classical mechanics, the identification and resolution of which led to relativity theory (Parts II, IIIA, IIIB).

According to the structured-time interpretation of q.m., the problem is once again with the notion of time in classical mechanics. The particular shortcoming in question is the simplistic concept of linear time used in classical mechanics. At the turn of this century, Poincaré's analysis of the deficiencies in the Newtonian concept of time, particularly the concept of simultaneity and equal intervals of time, led to the origin of the special theory of relativity. However, according to the structured-time interpretation of q.m., this analysis did not go far enough, for it assumed that time was like a line, by assuming the existence of *intervals* of time, and dealing only with the problem of the equality of the lengths of these intervals. According to the structured-time interpretation, the assumption of time as a line (either a straight line or a timelike curve) needs to be given up.

### *4. 2  Overview of the argument*

As the argument is a little complex, we present an overview first. We argue below (§ 4.3) that a microphysically structured time follows as a consequence if one accepts small departures from strict time-asymmetry or a small *'tilt in the arrow of time'*. We recall that the hypothesis of a tilt in the arrow of time itself emerged naturally in the process of attempting to establish a physical direction of time, going from the entropy law (Part IV) to the electromagnetic field (Part VA) to electrodynamics (Part VB).

The link with q.m. is obtained as follows. The difference between classical mechanics and q.m. was related above to a difference of logic. Now, a microphysically structured time, too, entails a change of logic at the microphysical level: the logic of a structured time differs from the logic of linear time in somewhat the same way as the logic (required for the formal semantics) of a parallel computer program differs from the logic of a sequential program. What remains to be seen is whether the logic arising from a microphysically structured time corresponds to a quantum logic.

The above claims may be represented schematically.

time-asymmetry in classical mechanics $\longrightarrow$ tilt in the arrow of time $\longrightarrow$

structured time $\longrightarrow$ change of logic $\longrightarrow$ quantum logic $\longrightarrow$ q.m.

The slogan formulation of these claims would be: (1) 'Non-locality is unavoidable in classical mechanics', and (2) 'Non-local classical mechanics is quantum mechanics'.

*4. 3  From electrodynamics to structured time*

As a specific and empirically viable example of a tilt in the arrow of time let us consider direct-action electrodynamics with incomplete absorption. That is, we go back to the situation, considered in Part VB, where interactions are both retarded and advanced, but advanced interactions form only a small part, say 1 part in a trillion. 'Strict causality' or perfect time asymmetry has broken down, but the departures from the standard picture are quantitatively rather small.

The particle[16] equations of motion are now differential equations with mixed-type deviating arguments considered in Part VB. As pointed out there, such equations do not, in general, admit a unique solution, in either direction in time, even if the entire past (or future) history is prescribed. Typical solutions will *branch* and *collapse* both into the future and the past. The picture is somewhat similar to what one would get by taking paths, in the path integral formalism, literally as particle trajectories.

Accepting the existence of small amounts of advanced radiation as an empirical hypothesis, what are we to make of this non-uniqueness, and the branching and collapse of solutions? In the absence of any mathematical or physical criteria which enable one to select a particular solution, the structured-time interpretation grants equal validity to all the solutions.

Granting equal validity to all solutions means that *the state of the world is not uniquely specifiable*: the world has split into a number of equally real states.[17] The statement, 'the $x$-coordinate of the electron (at "time $t$") is $x_0$' cannot any longer be regarded as either true or false: its truth or falsity depends upon the particular branch of the solution we choose. Let us refer to this state of affairs, involving intrinsic non-uniqueness, and the branching of solutions as 'branching time'.[18]

It is impossible to describe branching time, or an 'ambiguous reality', with the usual logic (of linear time), where a given statement must be either true or false, where the $x$-coordinate of an electron (at time $t$) either definitely is, or is not, $x_0$, *regardless of our state of knowledge*. A new logic is needed which takes into account the generic breakdown of uniqueness of solutions, and the resulting breakdown of determinism, at least at the microphysical level.

*4. 3. 1  The analogy with CSP.*  The salient features of this new logic are not difficult to grasp. For those familiar with parallel computing, there is an interesting and complete analogy with the branching of processes in Hoare's language of communicating sequential processes (CSP), and its practical implementation in OCCAM. Unlike the case of the transactional interpretation of q.m., the analogy here is incidental rather than essential, and operates at the deeper level of the temporal logic required for the formal semantics of Hoare's CSP language.[19] Consider an OCCAM program of the following type.

```
        PAR                         - - do indented processes in parallel
          PAR i=0 FOR n             - - Proc 1: replicate in parallel
            SEQ                     - - do in sequence
              A(i)                  - - some process which generates B(i)
              Chan [i] ! B(i)       - - output B(i) to Chan (i)
          ALT i=0 FOR n             - - Proc 2: do indeterministically
            Chan [i] ? B            - - input from Chan (i) to B
```

The formal indeterminism of the ALT construct should not be confused with 'randomness': the ALT construct does *not* assert, 'select one of the *B*(*i*) at random'. *B* is not a classical random variable, since it is impossible, *in principle*, to specify uniquely[20] the value of the number *B*, or the final state of the program (assuming that the numbers *B*(*i*) are distinct). The PAR construct corresponds to branching: the *n* processes evolve in parallel. The ALT corresponds to indeterministic collapse: one of the *n* values is selected indeterministically.

*4. 3. 2 Past and present contingents.*  We note that, with the hypotheses of a tilting arrow of time, time branches into both the future and the past, as compared to the everyday picture of time as future-branching but past-linear.[21] In the usual picture (built into the grammatical structure of Indo-European languages), which we take seriously at the level of everyday life (and in the belief that one is 'free' to perform any experiment), the future is uncertain or contingent, but the belief is that the past is decided (even if we do not *know* all about it). But with a tilt in the arrow of time not only the future, but the past and the present also become contingent.[22]

*4. 3. 3 The logic of structured time.*  The simplest logic that describes such a structured time is a quasi truth-functional (q.t.f.) logic, considered in Part I. In this logic there are precisely two branches, corresponding to a world with two possibilities (at any 'instant'). By considering collections of such 'worlds' one may model a situation with many possibilities.

We now explain how such a logic naturally gives rise to a quantum logic.

*4. 4  From structured time to quantum logic*

*4. 4. 1 Statements.*  To model the concept of a world with many possibilities, in a logical setting, each 'branch' or possibility of the real world, at any 'instant', is modeled by the Wittgensteinian concept of a world as 'all that is the case'. In other words, a world, or rather a truth-functional world, is specified by a collection of statements deemed to be true. Intuitively, the statements declared to be true are tautologies, and the empirical facts valid at that 'point of time'.

The primary model of empirical facts being that of classical mechanics, the basic statements, *p*, are taken to be of the type

$$p \equiv m \in E, \tag{14}$$

where *m* is a classical dynamical variable, and $E \in B_R$ is a Borel subset of *R*. Reference to (3) and (4) shows that the probabilities of interest in q.m. are the probabilities of precisely such statements. As explained earlier, the 'quantum' nature of these probabilities is decided by the logic used to compound these basic statements.

The set of all propositions[23] of type (14) will be denoted by **P**. This seems simple enough, except that there is a subtle but crucial difference right here. As Van Fraassen[24] points out, 'The description of the set of propositions is usually quite abstract, the relation to q.m. being left to the imagination of the reader.'

The point is that it is problematic to assign *truth-values* T, F or 0, 1 or 'yes', 'no' to sentences of the type (14). This is straightforward in classical mechanics, but in quantum mechanics only *probabilities* of statements of type (14) are meaningful. In the BN approach, the elements of **P** are related to true-false *experiments*, and in the Jauch-Piron approach to yes-no *questions*. In all such motivations, the measurement process enters as a logically primitive notion, even though all the controversy in q.m. centres around the measurement process.

In the present approach, elements of **P** correspond to well-formed statements of classical mechanics, which admit a truth-value *if* a classical state is prescribed. The lattice theory approach has been regarded as suspect precisely because, for quantum-mechanically incompatible statements *p*, *q*, experiments for *p* and *q* interfere, so that $p \wedge q$ is not physically *meaningful*. So the objection to a lattice structure is no longer valid with the present approach. The set of propositions may form a lattice, even if one does not expect this for the set of *measurable* propositions.

*4. 4. 2 Truth-functional worlds.* A truth-functional world $\alpha$ is, then, a collection of statements *p*, such that (i) every tautology is in $\alpha$, (ii) conjuncts ($p \wedge q$), disjuncts ($p \vee `q$), and logical consequences of elements *p*, *q* of $\alpha$ are in $\alpha$, and (iii) for any *p*, either *p* is in $\alpha$ or its negation ($\sim p$) is in $\alpha$. Since the only empirical facts of concern are those of classical mechanics, a truth-functional world may be thought of as specified by a point in classical phase space.

*4. 4. 3 Quasi truth-functional worlds.* In a truth-functional world every statement is either true or false. In a world admitting contingents, or in a world in which time branches, there must be at least two possibilities or branches. The simplest logical model of such a world is, therefore, a **quasi truth-functional world** $\omega = \{\alpha, \beta\}$, an unordered pair of truth functional worlds $\alpha$, $\beta$. In the structured-time interpretation, a q.t.f. world is regarded as the basic element of (microphysical) reality.

*4. 4. 4 Modalities.* In a q.t.f. world, modalities (or the notions of 'possibility' and 'necessity') may be formally defined. A statement *p*, which is true in both $\alpha$ and $\beta$, is **necessary** (or necessarily true) in $\omega = \{\alpha, \beta\}$, denoted by $\omega \in |\Box p|$. A statement *p* which is true in at least one of $\alpha$, $\beta$ is called **possible** (or possibly true) in $\omega$, denoted by $\omega \in |\Diamond p|$. A statement which is false in both $\alpha$, $\beta$ is, of course, necessarily false in $\omega$.

*4. 4. 5 Irreducible contingents and admissible worlds.* A q.t.f. world represents a snapshot of a world with two possibilities. Since there is only an inconsequential difference between a q.t.f. world $\omega = \{\alpha, \alpha\}$ and the truth functional world $\alpha$, such worlds are ruled out by deeming them not **admissible.** The precise criteria for admissibility will be considered later: they have to do with preserving a minimum or irreducible contingency. The impossible world 0, in which every statement is true, and the true world 1, in which only tautologies are true, will be deemed, by convention, to be admissible.

*4. 4. 6 States.* The collection of all admissible q.t.f. worlds is the (logical) **universe** $U$ (and a q.t.f. world will by default be admissible from now on). A collection of admissible q.t.f. worlds is a **state**.[25] A state, thus, represents a snapshot of a world in which there are many possibilities, but the basic units of which are also irreducibly contingent. Using the correspondence of truth-functional worlds with points in phase space, a state may be identified with a swarm of points in classical phase space, with the proviso that the 'real' state of the system is *not* represented by any single point. A state represents the real world at an instant of time, or rather an instant of time may be defined by a state.

It will become clear later on that the connection between states and probability measures on the set of propositions is the following. Each q.t.f. world $\omega$ induces a probability measure $P_\omega$ defined on **P** as follows.

$$P_\omega(p) = \begin{cases} 1 & \text{if} & \omega \in |\Box p| \\ 1/2 & \text{if} & \omega \in |\Diamond p| \\ 0 & otherwise \end{cases}$$

(15)

*4. 4. 7 Quantum measurements, access relations, and selection functions.* The idea is to model the process of quantum measurement as follows. A measurement can be carried out only if there is some certainty: it is pointless to speak of *the* length of a moving earthworm. The situation is worse if there are several possibilities at *the same 'instant of time'*, as happens with structured time. In a world with many possibilities one can measure only that which is certain. That is, a dynamical variable can be measured only in a world in which it necessarily assumes a certain value.

Thus, the measurement process for a given dynamical variable, $m$, must, by definition, transform the given world (or state) into one in which a statement of the type (14) is necessary. Therefore, the process of measurement, conditioned on a given statement, (discontinuously) 'selects' the 'closest' 'accessible' world in which the statement is necessary. We have already defined 'necessary', and we now formalize the terms in quotes.

Accessibility is modeled by means of a relation. An **access relation** R is defined between two q.t.f. worlds, $\omega_1$ and $\omega_2$, by requiring that $\omega_1$ R $\omega_2$ if and only if $\omega_1$ and $\omega_2$ have a truth-functional world in common, $\omega_1 \cap \omega_2 \neq \varnothing$. The impossible world will be deemed, by convention, to be accessible from any world. We observe that this access relation is not transitive.

'Selection' is modeled by a selection function. A **selection function**, conditioned on statement $p$, is a map between q.t.f. worlds such that (i) $f_p(\omega)$ is accessible from $\omega$, (ii) $p$ is necessary in $f_p(\omega)$, and (iii) if $p$ is necessary in $\omega$, then $f_p(\omega) = \omega$. More formally, a map $f{:}\mathbf{P}{\times}U{\to}U$, such that $f(p, \cdot)$ satisfies the above properties, for each $p$, is called a selection function.

The process of measurement is specified by a selection function,[26] which is expected to 'select' the 'closest' accessible q.t.f. world in which $p$ is necessary. In a general logical setting, defining 'closest' is known to create difficulties[27]: if Bizet and Verdi are friends, is the closest world one in which they are both French or both Italian? In the present case, remembering that a truth-functional world may be specified by a point in classical phase space, it is possible, though not essential, to define 'closest' as closest in the sense of distance in classical phase space, subject to admissibility (or the exclusion of a certain 'minimum' phase space volume in order to ensure that contingency is irreducible). The actual existence of a selection function is demonstrated later on.

*4. 4. 8 Incompatibility, joint measurements and repeated measurements.* In the above view of measurement, two statements $p$, $q$, are deemed to be **incompatible** if there is no q.t.f. world, hence no state, in which they are jointly necessary, but $q$ is possible in some worlds in which $p$ is necessary and vice versa. It is impossible, in principle, to jointly measure two incompatible statements: a measurement of $p$ results in a world in which there is no definite information about $q$. It may be easily checked that, in terms of selection functions, this corresponds to a failure of commutativity: $f_p \circ f_q \neq f_q \circ f_p$. That is, in terms of repeated measurements, for incompatible statements $p$, $q$, a measurement of $p$ followed by a measurement of $q$ does not, in general, result in the same world as a measurement of $q$ followed by a measurement of $p$.

*4. 4. 9 Measurability.* If $p$, $q$ are incompatible statements, then $p \wedge q$ is a statement which is possible in some worlds but necessary nowhere. Such a statement is not open to measurement. We now define the **quasi-measurable** statements as those $p$ for which, given a world $\omega$ in which $p$ is possible, it is *always* possible to select a (unique) non-impossible world $f_p(\omega) \neq 0$, in which $p$ is necessary. A proposition $p$ is regarded as **measurable** if both $p$ and its negation $\sim p$ are quasi-measurable. The set of all measurable propositions will be denoted by **M**. Formally,

$$\overline{M} = \{p \in \boldsymbol{P} \,|\, f_p(\omega) \neq 0, \text{ for all } \omega \in |\Diamond p|\},$$

$$\mathbf{M} = \{p \in \boldsymbol{P} \,|\, p \in \overline{M} \wedge \sim p \in \overline{M}\} \quad . \tag{16}$$

*4. 4. 10 'And', 'or'; 'if', 'not'.* As observed earlier, the origin of the difference between classical and quantum probabilities, as also many of the differences between the two formalisms, can be traced to the fact that 'and' and 'or' between statements must be interpreted differently in q.m.. As also seen earlier, it is possible and more convenient to work with a new type of 'if' or a new type of order relation ('implicate order') which we now introduce, using the concepts developed above.

It is, first of all, clear that in a world admitting contingents, a new type of 'if' is required. What would one say about the truth of $p \Rightarrow q$ in a world in which $p$ is *possible*, but $q$ is false? One may adopt different positions. Two simple possibilities are the following:

$$\Box p \Rightarrow \Box q, \tag{17}$$

$$\Box(p \Rightarrow q). \tag{18}$$

The first says that 'if $p$ then $q$' is deemed true exactly if whenever $p$ is necessarily true so is $q$. If $p$ is only possible, nothing is implied about $q$. The second (strong implication) says that $p \Rightarrow q$ must necessarily be true, and is the stronger formulation: this implication is false if there is a world in which $p$ is possible but $q$ is false. A variety of other formulations exist.[28]

*4. 4. 11 Preliminary order relation.* The preliminary order is defined on the set **P** of all propositions[29] $p$ of the form (14), using the selection function introduced above. We introduce a proposition $p » q$ as follows:

$$\omega \in |p » q| \equiv f_q \circ f_p(\omega) = f_p(\omega). \tag{19}$$

This says that $p » q$ is true in a world $\omega$ exactly if the closest world in which $p$ is necessary is also a world in which $q$ is necessary. One now gets an order relation $p > q$ by the proposition $\Box_U (p » q)$,

$$p > q \equiv \Box_U (p » q), \tag{20}$$

i.e., by demanding that $p » q$ be true in every admissible world.

It is clear that the 'if' ('counterfactual implication') or order relation defined by (20) is weaker than that defined by (18). One of the achievements of the theory is the rather easy proof[30] that, for *measurable* propositions, the order relation, defined above in terms of repeated measurements, may be characterized in terms of strong implication:

$$p > q \Leftrightarrow \Box_U (\Box (p \Rightarrow q)), \quad p, q \in \mathbf{M}. \tag{21}$$

This order relation is not, in general, antisymmetric, for a proposition $p$ which is possible in some worlds but necessary nowhere satisfies both $p > 0$ and $0 > p$, without being identically false. But we may adopt the usual algebraic technique of identification and equivalence classes: we regard $p$ and $q$ as **weakly equivalent**, $p \approx q$, if both $p > q$ and $q > p$ hold, in order to obtain an order relation on the set of equivalence classes $\mathbf{P}/\approx$. This is not algebraic juggling — the measurement process cannot distinguish between weakly equivalent statements. For $p, q \in \mathbf{M}$, of course, weak equivalence is the same thing as usual logical equivalence.

*4. 4. 12 Orthogonality, compatibility and order relation.* Before defining the final '*if*', we need one last concept. Two propositions $p$ and $q$ are regarded as **orthogonal**, written $p \perp q$, if $p > \sim q$ holds,

$$p \perp q \equiv p > \sim q. \tag{22}$$

Using orthogonality, we can finally define an order relation among measurable propositions as follows: $p \gg q$ if and only if there exists a measurable proposition $s$ orthogonal to $p$ such that $q = p \vee s$:

$$p \gg q \equiv \exists s \in \mathbf{M}, \ s \perp p \ \text{and} \ q = p \vee s. \tag{23}$$

The earlier order relation seemed fairly natural, but what is the significance of this new relation? It may be shown[31] that this order relation is equivalent to (i) the earlier order relation $p > q$ **together** with (ii) compatibility, or commutativity of the selection functions. Algebraically, the advantage of defining the order relation in this way is that orthomodularity is an immediate consequence.

*4. 4. 13 Main result.* The main result is as follows.
**Theorem.**[32] The set of all measurable propositions, **M**, ordered by the '*if*' defined by (23), $>>$, and complemented by the usual '*not*', $\sim$, is a quantum logic which is also atomic.

(The terminology used in reference 15 and the current terminology may be related as follows. Given a countable collection of measurable propositions $p_i \in \mathbf{M}$, which are mutually orthogonal, $p_i \perp p_j$ for $i \neq j$, the measurability of $\vee_\iota p_i$ follows because not more than one $p_i$ may be true in a truth-functional world. For the same reason, $P_\omega$, defined by (15), is countably additive, hence a probability measure. Hence, using (23), if $p >> q$ then $P_\omega(q) = P_\omega(p) + P_\omega(s)$. If $s \neq 0$, then there exists an $\omega$ for which $P_\omega(s) \neq 0$; for this $\omega$, $P_\omega(p) < P_\omega(q)$. That is, there exists an order-determining set of states.)

We can now summarize. Starting from a purely classical hypothesis (the existence of small amounts of advanced radiation), we have arrived at the basic minimum algebraic structure required for q.m., and particularly for the mysterious behaviour of quantum probabilities.

There is undoubtedly a gap between the slogan formulation and what has been achieved above. Some questions about semi-modularity, irreducibility, and the lattice structure are considered below. The more serious deductive gap pertains to the irreducibility of contingents assumed below, in the criteria of admissibility. The examples in Part VB only showed that contingents are possible; below we make the inductive generalization that they are necessary. To put matters crudely, the physically plausible assumption is that the 'degree' of contingency is proportional to the amount of advanced interactions.

In the absence of a well-developed mathematical theory of functional differential equations, one may argue for the plausibility of this assumption as follows. Consider the motion of a macrophysical object such as a rocket. Emission of a single advanced photon from the rocket (= absorption of a single retarded photon) is closely analogous to the usual statistical perturbation (even though this perturbation could not, in principle, have been determined from an exact knowledge of the entire past history of the rocket and the enveloping universe). In this situation, one would expect possible deviations from the classical trajectory to be small, just because these would be proportional to the strength of the perturbation, i.e., the amount of advanced interactions, which has been assumed to be very small.

*4. 4. 14 Example of failure of the distributive law.* With the above order relation, the distributive law, of course, fails. Let us call a q.t.f. world in which $p$ is necessary a $p$-world. Choose measurable propositions $p, q, s$, such that no $p$-world or $q$-world is an $s$-world. Then $s \wedge p \approx 0$, $s \wedge q \approx 0$, but $s \wedge (p \vee q) \neq 0$, if $p$ is true in one branch of an $s$-world while $q$ is true in the other branch.

The failure of the distributive law arises through a 'squeezing in' of possibilities: the necessity of $p \vee q$ does not imply the necessity of $p$ or the necessity of $q$, for $p$ may be true in one branch while $q$ is true in the other. In the two-slit experiment, in the case of the diffraction pattern, it is necessary that the electron came through slit A or slit B. In the case of the Gaussians, it is *necessary* that the particle came through slit A or it is *necessary* that the particle came through slit B. The two statements are not equivalent.

The possibility that has got 'squeezed in' is not that the electron split into two, with one part going through slit A and the other part going through slit B. Rather, the real world has two components: in each the electron came through one slit. On this interpretation, the diffraction pattern is an empirical consequence of a contingent past.

*4. 4. 15 The Lattice structure.* We had remarked earlier that alternative approaches to quantum axiomatics tend to reject the lattice structure, but that such a structure is possible in the present approach, because the propositions are not motivated either in terms of extant quantum theory or using some logically primitive notion of the measurement process. Thus, there is no question of the physical interpretation of the lattice structure. The B-N question, 'What is the physical significance of $\wedge, \vee$ ?', is to be seen as a wrong question dating back to the days when physics was regarded as no more than an axiomatic theory together with an interpretation.

Rather, the lattice is to be seen as a technical convenience (like Borel sets), and obtained by enlarging, if necessary, the set of measurable propositions. Thus, the question is 'What is the most economical enlargement of **M** which will give a lattice?'.

Using the work of Pool[33] one may proceed as follows. Each $p \in \mathbf{M}$ may be identified with the function $f_p$. We let

$$L = \{f_{p_1} \circ f_{p_2} \circ \ldots \circ f_{p_n},\ p_1, p_2, \ldots, p_n \in \mathbf{M}\}. \tag{24}$$

$L$ contains a copy $P(L)$ of the quantum logic **M**, which can be extended to a lattice if

$$f_{p_1} \circ f_{p_2} \circ \ldots \circ f_{p_n} = f_{q_1} \circ f_{q_2} \circ \ldots \circ f_{q_n} \implies$$

$$f_{p_n} \circ \ldots \circ f_{p_2} \circ f_{p_1} = f_{q_n} \circ \ldots \circ f_{q_2} \circ f_{q_1}. \tag{25}$$

It is possible to interpret (24) as repeated measurements regarded as instantaneous in a coarse treatment, and (25) in terms of the micro time symmetry of repeated measurements; from the present point of view, however, the more important question is whether any of the selection functions defined below satisfy (25).

*4. 4. 16 Admissibility.* We pointed out earlier that q.t.f. worlds of the type $\omega=\{\alpha, \alpha\}$ cannot be deemed to be admissible. On the other hand, permitting q.t.f. worlds of the type $\omega=\{\alpha, \beta\}$, where $\alpha, \beta$ are arbitrarily close (in the sense that the corresponding points in phase space are arbitrarily close) would permit contingency to disappear by the back door.

Contingency may be maintained only if $\beta$ is excluded from a certain 'finite' (as opposed to infinitesimal) region around $\alpha$. To do this invariantly, one must exclude a 'finite' volume around $\alpha$. In the one-dimensional case, if $\alpha$ is specified by the phase space coordinates $(p_0, q_0)$, and $\beta = (p, q)$, then $\omega=\{\alpha, \beta\}$ is admissible, provided $|p - p_0|\ |q - q_0| \geq 1$ (in appropriate units). The existence of incompatible propositions, assumed earlier, is an immediate consequence of this criterion.

Such a criterion is open to the charge of having introduced the uncertainty principle by the back door. However, the structured-time interpretation is based on a dynamical hypothesis regarding the nature of the equations of motion. So, the above criterion has the status, not of an absolute principle, but of a conjecture concerning the solutions of the equations of motion. (It may not be so easy to settle this conjecture, since that would amount to relating the value of Planck's constant to the amount of advanced interactions assumed to exist.)

*4. 4. 17 Example of a selection function.* Given the above criterion of admissibility, one may construct a selection function as follows. Let $m$ be a classical dynamical variable, $E \in \boldsymbol{B_R}$, and $p \equiv m \in E$ the proposition described by (14). Let $\omega = \{\alpha, \beta\}$ be a q.t.f. world, where $\alpha$ and $\beta$ are described by the same phase space coordinates as above. If $(p_0, q_0) \in m^{-1}(E)$, i.e., the statement $m \in E$ is true in $\alpha = (p_0, q_0)$, we let $f_p(\omega) = \omega' = \{\alpha, \beta'\}$, where the *subscript p* refers to the statement $m \in E$, and $\beta'$ is specified by the unique phase space point $(p_1, q_1)$, closest to $(p, q)$, which satisfies the criterion of admissibility and also $(p_1, q_1) \in m^{-1}(E)$. In case such a point fails to exist or fails to be unique, we let $f_p(\omega) = 0$.

This example of a selection function is not unique. Instead of making the 'minimum' change in $\omega$, we could allow the selection function to minimize contingency by selecting the world closest to $\beta$ from among the admissible worlds closest to $\alpha$. The particular selection function one chooses decides the nature of the measurement process.

*4. 5 Relation to other interpretations*

*4. 5. 1 The many-worlds interpretation.* Though it is the world, rather than the electron, which suffers from schizophrenia, the structured-time interpretation must be distinguished from the superficially similar Everett-Wheeler 'many-worlds' interpretation of q.m. The many-worlds interpretation takes quantum-mechanics seriously enough at the macrophysical level to set up a wavefunction for the universe, leading to questions about the observer being included in the wave-function or consciousness being included in physics but excluded from the wave-function. To account for the collapse of the wave function, in the 'many-worlds' interpretation, the world constantly splits into possibilities (for no earthly reason). These are entirely separate worlds, and one observer is aware of only one possibility. Moreover, that interpretation does not provide any route to the formalism of q.m..

*4. 5. 2 The transactional interpretation.* The transactional interpretation[34] of q.m. is also superficially similar, in that it uses the Wheeler-Feynman theory and the 'handshake' analogy of the transputer implementation of CSP to account for non-locality: a retarded 'offer wave' generates an advanced 'confirmation' wave, resulting in a giant handshake across spacetime. The interpretation is primarily concerned with this analogy, which it regards as analogous to Faraday's lines of force!

However, we have already pointed out (Part VB) the internal inconsistency of the Wheeler-Feynman theory. The transactional interpretation also shows a curious disregard for any empirical consequences of non-locality beyond the possibility of interactions at space-like separations in Aspect-type experiments.

*4. 5. 3 The modal interpretation.* Though the structured time approach is constructive, rather than axiomatic, it is very similar to the modal interpretation.[35] One of the crucial differences is the use of quasi truth-functional semantics which greatly clarifies certain obscurities in the modal interpretation.[36]

*4. 5. 4 The Copenhagen interpretation.* The Copenhagen interpretation is, in a way, closest to the structured-time interpretation. The difference lies in the fact that the structured-time interpretation begins with a dynamical hypothesis rather than abstract philosophical principles like 'complementarity'. Rather general connections[37] between 'equilibrium' (or indifference to the origin of time) and unitary evolution, together with the analogy to the path integral formalism, suggest that there is at least some hope of linking classical Hamiltonian evolution with Schrödinger evolution, within the structured-time interpretation.

## 5  Conclusions

Non-local classical mechanics admits qualitative features that have been regarded as typically quantum mechanical. Probabilities in a world admitting irreducible contingents manifest features characteristic of quantum probabilities: specifically, the microphysically structured time, resulting from a microphysical tilt in the arrow of time, forces a change of logic, resulting in a quantum logic.

## Notes and References

1. Though we do not enter into this question here, a structured time is also closer to the time of daily experience.
2. For those who have only vaguely heard of measure theory, the significance of Borel sets is that it always makes sense to speak of the measure of such a set with respect to a regular Borel measure! The significance of a (finite) regular Borel measure is that it corresponds (both ways) to an integral which will (i) integrate continuous functions, and (ii) provide a finite value for the integral of a continuous function which 'vanishes at infinity', i.e., is less than the proverbial epsilon outside a compact set. The formal definition of a Borel set in a topological space $X$ is that it is an element of $B_X$, the smallest σ-algebra which contains all open sets.
3. P. R. Halmos, *Introduction to Hilbert space and the Theory of Spectral Multiplicity*, Chelsea, New York, 1951. The whole book is devoted to the proof of this theorem.
4. From Dirac's classic treatise on *Quantum Mechanics*, as cited by J.S. Bell, *Speakable and Unspeakable in Quantum Mechanics* (Cambridge: University Press, 1987), p 40.
5. For the difficulties in extending von Neumann's collapse postulate to observables with continuous spectra, and for a new formulation see M.D. Srinivas, 'Collapse postulate for observables with continuous spectra', Commun. Math. Phys., **71**, 131-58 (1980).
6. E. P. Wigner, Phys. Rev., **40**, 749 (1932).
7. E. P. Wigner, in: *Perspectives in Quantum Theory*, W. Yourgrau and A. van der Merwe (eds), MIT Press, Cambridge, Mass., 1971.
8. The other possibility, pursued for instance in stochastic quantization, is to change the axiomatic basis of q.m., so that joint probability distributions do exist even for canonically conjugate variables.
9. For the measure-theoretic foundations of probability theory, as enunciated by Kolmogorov, see, for instance, P. Halmos, *Measure Theory*, D. Van Nostrand, New York, 1950. To change from sets to sentences, see, e.g., J. R. Lucas, *The Foundations of Probability*, Clarendon, Oxford, 1957.
10. The logico-algebraic approach was initiated by G. Birkhoff and J. von Neumann, Ann. Math., **37**, 823 (1936). For further developments of this theory, see V.S. Varadarajan, *Geometry of Quantum Theory*, Vol. 1, Van Nostrand, Princeton, N.J., 1968, and C. A. Hooker (ed) *The Logico-Algebraic Approach to Quantum Mechanics*, D. Reidel, Dordrecht, 1975.
11. A 'subspace' will always mean a *closed* linear manifold.
12. One has to be a little careful in applying the analogy to implication. The sentential connective $\Rightarrow$ should not be confused with the *relation* of implication. Thus, it makes sense to speak of $A \Rightarrow (B \Rightarrow A)$, but not of $a \leq (b \leq a)$.
13. See G. Kalmbach, *Hilbert Lattices*, World Scientific, Singapore, 1987, for a counter-example, which however, does not apply to probability measures.
14. Those interested in probing further may like to be discouraged by the text of S. Lang, *Algebra*, Addison-Wesley, Reading, Mass., 1965.
15. C. K. Raju, 'Quantum mechanics and the microphysical structure of time' [1990]. In: R. Nair (ed) *In Search of Quantum Reality* (to appear).
16. This is true directly for charged particles, and indirectly, at least, for all particles which interact in any way with charged particles.
17. The whole debate over 'hidden variables' is related to the metaphysical assumption (of linear time) that 'reality is unambiguous'. This assumption breaks down with branching time.
18. The terminology derives from temporal logic.
19. See, e.g., K.R. Apt (ed), *Logics and Models of Concurrent Systems*, Springer, Berlin, 1984, NATO ASI Series in Computers and Systems Science, Vol. 13. The moral of the story is that one should link non-locality to CSP rather than ESP!
20. The usual definition of a random variable as a measurable *function* assumes the possibility of such an in-principle specification.
21. N. Rescher and A. Urquhart, *Temporal Logic* (Wien: Springer, 1971).
22. This does not lead to any paradoxes, as discussed in detail in Chapter VB. In terms of the analogy to the formally similar but more pragmatic CSP, the ALT construct of OCCAM is non-deterministic, and the contingent 'present' is what accounts for the current practical difficulty in constructing a debugger for parallel processors. Incidentally, in India, this difficulty has been currently resolved by abandoning CSP and adopting a queuing model in the PARAS system software developed by C-DAC for use with parallel FORTRAN, and C, which have some constructs resembling the PAR and ALT of OCCAM. The empirical significance of a contingent past is discussed later on.

23. Traditionally, a proposition is what a sentence expresses or means. Formally, 'meaning' may be deemed to have been assigned to a sentence *p* by specifying the set of 'worlds', |*p*|, in which the sentence is true, so that one knows when the sentence is true and when it is false. Since we shall be using different notions of 'world', we should have, strictly, distinguished between statements, truth-functional propositions and quasi truth-functional propositions. We shall, however, not retain this difference explicitly, in the hope that the type of entity being considered would be clear from the context.
24. Bas C. Van Fraassen, in: Hooker (ed) *Contemporary Research in Foundations and Philosophy of Quantum Theory*, D. Reidel, Dordrecht, 1973, p 84.
25. Strictly speaking, a state should be identified with a convex combination of the probability measures induced by each q.t.f. world in the collection. Thus, to each q.t.f. world ω in a state, one must also assign a number $t(\omega)$, $0 < t(\omega) < 1$, such that $\Sigma\, t(\omega) = 1$, the sum being taken over all q.t.f. worlds in the state.
26. The selection function is, in fact, a projection: $f_p \circ f_p = f_p$, and could, under the circumstance (25), quite literally be identified with a projection operator in Hilbert space.
27. W. L. Harper, R. Stalnaker, and G. Pearce (eds), *Ifs*, D. Reidel, Dordrecht, 1981.
28. See Harper et al, Ref. 27.
29. I will use statements and propositions interchangeably; see note 23.
30. Ref. 15, Theorems 1, and 2.
31. Ref. 15, Theorem 5, equivalence of (i) and (v).
32. Ref. 15.
33. J.C.T. Pool, (reproduced in Hooker, Ref. 24 above). In the next paper in Hooker (ed), Pool also considers the possible significance of semi-modularity.
34. J. G. Cramers, Rev. Mod. Phys., **58**, 647 (1986).
35. See van Fraassen, Ref. 24.
36. As a specific example, consider van Fraassen's assumption (2), in Ref. 24, corresponding to Pool's Axiom 1.3, in Ref. 33. This says that $\Box p \Rightarrow \Box q$ implies $\Diamond p \Rightarrow \Diamond q$. That is, if the necessity of *p* implies the necessity
17. The whole debate over 'hidden variables' is related to the metaphysical assumption (of linear time) that 'reality is unambiguous'. This assumption breaks down with branching time.
18. The terminology derives from temporal logic.
19. See, e.g., K.R. Apt (ed), *Logics and Models of Concurrent Systems*, Springer, Berlin, 1984, NATO ASI Series in Computers and Systems Science, Vol. 13. The moral of the story is that one should link non-locality to CSP rather than ESP!
20. The usual definition of a random variable as a measurable *function* assumes the possibility of such an in-principle specification.
21. N. Rescher and A. Urquhart, *Temporal Logic* (Wien: Springer, 1971).
22. This does not lead to any paradoxes, as discussed in detail in Chapter VB. In terms of the analogy to the formally similar but more pragmatic CSP, the ALT construct of OCCAM is non-deterministic, and the contingent 'present' is what accounts for the current practical difficulty in constructing a debugger for parallel processors. Incidentally, in India, this difficulty has been currently resolved by abandoning CSP and adopting a queuing model in the PARAS system software developed by C-DAC for use with parallel FORTRAN, and C, which have some constructs resembling the PAR and ALT of OCCAM. The empirical significance of a contingent past is discussed later on.